\documentclass[letterpaper]{article}
\usepackage{aaai17}
\usepackage[sort,numbers]{natbib}

\usepackage{times}
\usepackage{latexsym}
\usepackage{asonimacros}
\usepackage{multirow}
\usepackage{epsfig}
\usepackage{algorithm, algorithmic}

\title{RIPML: A Restricted Isometry Property based Approach to Multilabel Learning}

\author{Akshay Soni \\ Yahoo! Research, Sunnyvale \\ akshaysoni@yahoo-inc.com \And Yashar Mehdad \\ Airbnb, San Francisco \\yashar.mehdad@airbnb.com }

\begin{document}
\maketitle

\begin{abstract}
The multilabel learning problem with large number of labels, features, and data-points has generated a tremendous interest recently. A recurring theme of these problems is that only a few labels are active in any given datapoint as compared to the total number of labels. However, only a small number of existing work take direct advantage of this inherent extreme sparsity in the label space. By the virtue of \emph{Restricted Isometry Property} (RIP), satisfied by many random ensembles, we propose a novel procedure for multilabel learning known as RIPML. During the training phase, in RIPML, labels are projected onto a random low-dimensional subspace followed by solving a least-square problem in this subspace. Inference is done by a k-nearest neighbor (kNN) based approach. We demonstrate the effectiveness of RIPML by conducting extensive simulations and comparing results with the state-of-the-art linear dimensionality reduction based approaches. 
\end{abstract}

\section{Introduction}
The task of multilabel learning is to predict a small set of labels associated with each datapoint out of all possible labels. Interest in these problems with large number of labels, features, and data-points has risen due to the applications in the area of image/video annotation \cite{Qi:2007}, bioinformatics where a gene has to be associated with different functions \cite{Barutcuoglu:2006}, and entity recommendation for documents and images on a web-scale \cite{Katakis:2008,Boutell:2004}. Modern applications of multilabel learning are motivated by recommendation and ranking problems; for instance, in \cite{Agrawal:2013} each search engine query is treated as a label and the task is to get the most relevant queries to a given webpage.  Specific to Natural Language Processing (NLP), developing highly scalable approaches for multilabel text categorization is an important task for variety of applications such as relevance modeling, entity recommendation, topic labeling and relation extraction.

Recently, dimensionality reduction based approaches have gained popularity, for example, by using Compressive Sensing (CS) \cite{Hsu:2009,Kapoor:2012} and the state-of-the-art Low Rank Empirical Risk Minimization (LEML) algorithm \cite{Yu:2014}. There has also been advances made in non-linear dimensionality reduction based approaches such as the X1 algorithm \cite{Bhatia:2015}. These algorithms, even though being conceptually simple, are still computationally heavy. For instance, Compressive Sensing based approach has a very simple dimensionality reduction procedure based on random projections, but require to solve a sparse reconstruction problem during prediction which is the bottleneck. 

To address these issues, we propose a novel approach that leverages the advantages of both Compressive Sensing and the non-linear X1 algorithm. RIPML benefits from a simple random projection based dimensionality reduction technique during training as in Compressive Sensing and then use a kNN based approach during inference as recently proposed in the X1 algorithm \cite{Bhatia:2015}. The proposed approach is based on the fact that the number of active labels associated with a datapoint is significantly smaller than the total number of labels, making the label vectors sparse. During training, we exploit this inherent sparsity in the label space by using random projections as a means to reduce the dimensionality of the label space. By the virtue of Restricted Isometry Property (RIP), satisfied by many random ensembles, the distances between the sparse label vectors are approximately preserved in the projected low-dimensional space as well. Given the training feature vectors, we then solve a least-squares problem to predict the low-dimensional label vectors.

During inference, for a new datapoint, we use the output of the least-square problem to estimate the corresponding low-dimensional label vector and then use kNN in the low-dimensional label space to find the $k$-closest label vectors. In this way, the labels that occur many times in these $k$-closest label vectors then become the estimated labels for this new datapoint. However, as noted by authors in \cite{Bhatia:2015}, kNN is known to be slow if the search for nearest neighbors involve large number of data points which is generally the case. We then leverage the solution provided in \cite{Bhatia:2015} and cluster the training data into multiple clusters and apply RIPML to each cluster separately.  



\subsection{Related Work}
The main advantages of embedding based methods is their simplicity, ease of implementation, strong theoretical foundations, the ability to handle label correlations, the ability to adapt to online and incremental scenarios, and the ability to work in a language/domain ignorant manner. The idea of Compressive Sensing based approaches \cite{Hsu:2009,Kapoor:2012} is to project the high-dimensional label vector into a smaller random-subspace and then solve a sparse recovery problem in this low-dimensional space.  The state-of-the-art (LEML) algorithm \cite{Yu:2014} leverages the low-rank of label matrix to learn the projection matrix and the back-projection matrix in order to estimate the label vectors by solving a single unified optimization problem.

The X1 algorithm \cite{Bhatia:2015} builds on the assertion that the critical assumption made by most dimensionality reduction based methods that the training label matrix is low-rank is violated in almost all the real world applications. The authors propose a locally non-linear embedding technique to reduce the dimension of the label vectors while approximately preserving the distances between them. Prediction is done by using kNN in this low-dimensional space over the training data. 



\section{RIPML}
\subsection{Background}
In order to formulate the problem and present our approach, we first note down the definition of RIP and few matrices which satisfy this property.\\

\noindent \textbf{Definition:}  \emph{A matrix $\bPhi \in \RR^{m \times n}$ satisfy the $(k, \delta)$-RIP for $\delta \in (0,1)$, if
\begin{equation}
(1-\delta)\|\bx\|_2^2 \leq \|\bPhi \bx\|_2^2 \leq (1+\delta)\|\bx\|_2^2 
\end{equation}
for all $k$-sparse vectors $\bx \in \RR^n$}.

While it is difficult to construct deterministic matrices which satisfy RIP, the best known guarantees arise from the random matrix theory. For example, following random ensembles satisfy RIP with high probability \cite{Rudelson:2006}
\begin{itemize}
\item Gaussian matrix whose entries are i.i.d. $\cN(0, 1/m)$ i.e. distributed normally with variance $1/\sqrt{m}$ for $m = \cO(k\log(n/k))$ 
\item Bernoulli matrix with i.i.d. entries over $\{\pm 1/\sqrt{m}\}$ with $m = \cO(k\log(n/k))$
\end{itemize}
Note that if $n$ is large and $k$ is very small then we only need $m \ll n$ to satisfy RIP, giving a very low-dimensional distance preserving embedding. If a matrix $\bPhi$ satisfy $(2k, \delta)$-RIP, then for all $k$-sparse vectors $\bx$ and $\by$, we have
\begin{equation*}
(1-\delta)\|\bx - \by\|_2^2 \leq \|\bPhi (\bx - \by)\|_2^2 \leq (1+\delta)\|\bx - \by\|_2^2 
\end{equation*}
which essentially means that the distance between the projected vectors $\bPhi \bx$  and $\bPhi \by $ is close to the distance between the original vectors $\bx$ and $\by$. This distance preserving property of random projections is at the core of RIPML.

It is to be noted that the classical Johnson and Lindenstrauss Lemma  \cite{Dasgupta:2003:JL} shows that any set of points can be embedded in a lower-dimensional space while preserving the distances between them. RIP  specializes that result and proves that some random ensembles indeed have this property, and can take advantage of the underlying sparsity to find a space of $\cO(k\log(n/k))$  dimension to embed $n$ points that are $k$ sparse.

As noted above, there are many different random ensembles which satisfy RIP, but in this paper we report experimental results using Gaussian ensembles only.  



\subsection{Algorithm}
Training data is of the form $\{(\bx_i, \by_i), i = 1,2,\dots,N\}$, where $\bx_i \in \RR^d$ is the feature vector, $\by_i \in \{0,1\}^{L}$ is the binary label vector and $L$ denotes the total number of labels. For $\ell \in [L]$\footnote{Here and in rest of the paper, for a non-negative integer $M$, the notation $[M]$ represents the set  $\{1,2,\dots,M\}$.}, $\by_i[\ell] = 1$ denotes that the $\ell^{\rm th}$ label is ``present" and $\by_i[\ell] = 0$ denotes otherwise. 

		
	

\begin{algorithm}[t]
\small
	\caption{RIPML: Inference}
	\label{alg:training}
	\begin{algorithmic} 
		\STATE \hspace{-1.4em} \textbf{Inputs:} Test point $\bx_{\rm new}$, no. of desired labels $p$, no. of nearest neighbors $k$, number of learners $F$, $\bZ$,  $\widehat{\bPsi}^f$ for $f \in [F]$, $\bY$
		\STATE  \textbf{Step 1: } For each $f \in [F]$ do:
		\STATE  \hspace{2em}\textbf{a) }$\bz_{\rm new}^f = \widehat{\bPsi}^f \bx_{\rm new} $
		\STATE  \hspace{2em}\textbf{b) }$\{i_1^f, i_2^f, \dots, i_k^f\} \leftarrow \text{kNN}(k)$ in $\bZ$ 
		\STATE  \textbf{Step 3: }$D = \frac{1}{Fk}\sum_{f = 1}^{F}\sum_{i = i_1^f}^{i_k^f}\by_i$
		\STATE  \textbf{Step 4: }$\widehat{\by}_{\text{new}} \leftarrow \text{Top}_p(D)$
		\STATE \hspace{-1.4em} \textbf{Output:} $\widehat{\by}_{\text{new}}$
	\end{algorithmic}
\end{algorithm}

\subsubsection{Training Procedure}
\noindent\textbf{Step 1 -- Label Vector Dimensionality Reduction:}
First, we project the training label vectors into a lower-dimensional space while approximately preserving the distances between them. This by the virtue of sparsity of label vectors is achieved by using a RIP satisfying matrix as a dimensionality reduction operator. That is, given a RIP satisfying matrix $\bPhi \in \RR^{m \times L}$, we get the low-dimensional label vectors as
\begin{equation}
\bz_i = \bPhi\frac{\by_i}{\|\by_i\|_2} =  \bPhi \tilde{\by}_i
\label{eqn:step1}
\end{equation}
where $\bz_i \in \RR^m$ is the low-dimensional representation of $\by_i$. Note that the above matrix-vector product can be efficiently calculated by just adding entries of each row of $\bPhi$ corresponding to the nonzero locations of $\by_i$ and then normalizing the result by the square root of number of nonzero entries in $\by_i$. If there are $s$-nonzeros in $\by_i$, the above product can be computed in $\cO(sm)$ operations rather then $\cO(mL)$ operations, required if the label vectors were dense.  Since we are operating under the assumption that $s \ll L$, the dimensionality reduction procedure adopted by us is efficient and fast. We normalize the label vectors in \eqref{eqn:step1} in order to work with the cosine similarity as distance metric. 

\noindent\textbf{Step 2 -- Least-Squares:} Given $(\bx_i, \bz_i)$ for $i \in [N]$, we want to learn a matrix $\bPsi \in \RR^{m \times d}$ such that $\bz_i \approx \bPsi\bx_i~~~\text{for~all}~ i \in [N]$. We propose to solve following least-square problem to learn $\bPsi$
\begin{equation}
\label{least_squares}
\widehat{\bPsi} = \mathrm{arg}~\underset{\bPsi}{\mathrm{min}}~\frac{1}{2}\sum_{i = 1}^N(\bz_i - \bPsi \bx_i)^2 + \lambda \|\bPsi\|_F^2
\end{equation}  
where $\lambda \geq 0$ is the regularization parameter which controls the Frobenius norm\footnote{For a matrix $\bX \in \RR^{m \times n},~~\|\bX\|_F^2 = \sum_{i,j}X_{ij}^2$} of the learned matrix. For reasonable feature dimension $d$, we can solve \eqref{least_squares} in closed form, and if solving in closed form is not an option, we can use optimization approaches like gradient descent to solve it iteratively. The overall output of the training procedure is $\bZ = [\bz_1, \bz_2, \dots, \bz_N] \in \RR^{m \times N}$ and $\widehat{\bPsi}$.

Since our approach is randomized by the choice of $\bPhi$, we can learn multiple models for different instances of $\bPhi$, and combine their predictions to produce a more accurate model. Let $F$ be the number of learners, then our training procedure gives us $\widehat{\bPsi}^f$ where $f \in [F]$. Unless stated otherwise $F = 5$ throughout the paper.

\subsubsection{Inference Procedure}
Given a new feature vector $\bx_{\rm new}$, we want to predict the labels associated with it. Given $\widehat{\bPsi}^f$ for $f \in [F]$, the following two steps are executed: \\

\noindent\textbf{Step 1 -- Get $\bz_{\rm new}^f \in \RR^m$:} $\bz_{\rm new}^f = \widehat{\bPsi}^f \bx_{\rm new}.$\\

\noindent\textbf{Step 2 -- Find kNN of $\bz_{\rm new}^f$:} Finds the indices of $k$ vectors from $\bZ$ which are closest to $\bz_{\rm new}^f$ in terms of squared distance. Say those indices are $i_1^{f}, i_2^{f}, \dots, i_k^{f}$. Then we compute the empirical label distribution as $\frac{1}{Fk}\sum_{f = 1}^{F}\sum_{i = i_1^f}^{i_k^f}\by_i$ out of which we can pick out the top-$p$ locations corresponding to highest values and give them as an estimate of the labels associated with $\bx_{\rm new}$.

We use vanilla kNN for the experiments in this paper, but this step can be made scalable and fast by using techniques such as Locality Sensitive Hashing \cite{Indyk:1998:knn}. For certain random ensembles like Gaussian, the locality-sensitive functions are already well-known \cite{Datar:2004:locality}. In order to keep the exposition simple, we make note of these approaches, but use simple kNN to do the experiments.  

\subsection{Scaling to Large Datasets} Even though our training procedure is simple and scalable, kNN can be slow for datasets with large number of data points which increases the testing time. In order to tackle large datasets, we first cluster the feature vectors into $C$ clusters using a simple procedure like KMeans. Then for each cluster $c$, we get the low-dimensional label vectors $\bZ^c$ (Training -- Step 1) and learn $\widehat{\bPsi}^c$ (Training -- Step 2).

For a new test feature vector, we first find its cluster membership by finding the cluster-center closest to it, and then apply our testing procedure by using $\bZ^c$ and $\widehat{\bPsi}^c$ for that cluster. 




\begin{table*}[t]
\small
\centering
\begin{tabular}{|c|c|c|c|c|c|c|c|}
\hline \bf Dataset & $d$ & avg. nnz($\bx$) & $L$ & avg. nnz($\by$)&Total Datapoints & Train ($N$) & Test\\ \hline
Bibtex & $1836$ &$68.74$& $159$ & $2.40$& $7395$ & $4880$ &$2515$\\ \hline
EURLex & $5000$ & $236.69$& $3993$ & $5.31$& $19314$ & $17383$&$1931$ \\ \hline
Delicious & $500$ & $18.17$& $983$ & $19.03$& $16091$ & $12910$ &$3181$\\ \hline
Chinese Relevance Modeling &  $100 - 400$& dense & $391$ & $1.02$ & $5011$ & $4511$ & $500$\\ \hline
Entity Recommendation & $400$ & dense & $359524$ & $32.55$ & $510539$ & $500539$ & $10000$\\
\hline
\end{tabular}
\caption{Statistics of different datasets used in this paper. Here, avg. nnz($\by$)  denotes the average number of labels per data-point. Similarly, avg. nnz($\bx$) denotes the average number of non-zero features per data-point.}
\label{table:datastats}
\end{table*}

\begin{figure*}[h]
\centering
\begin{tabular}{ccc}
\includegraphics[scale=0.35]{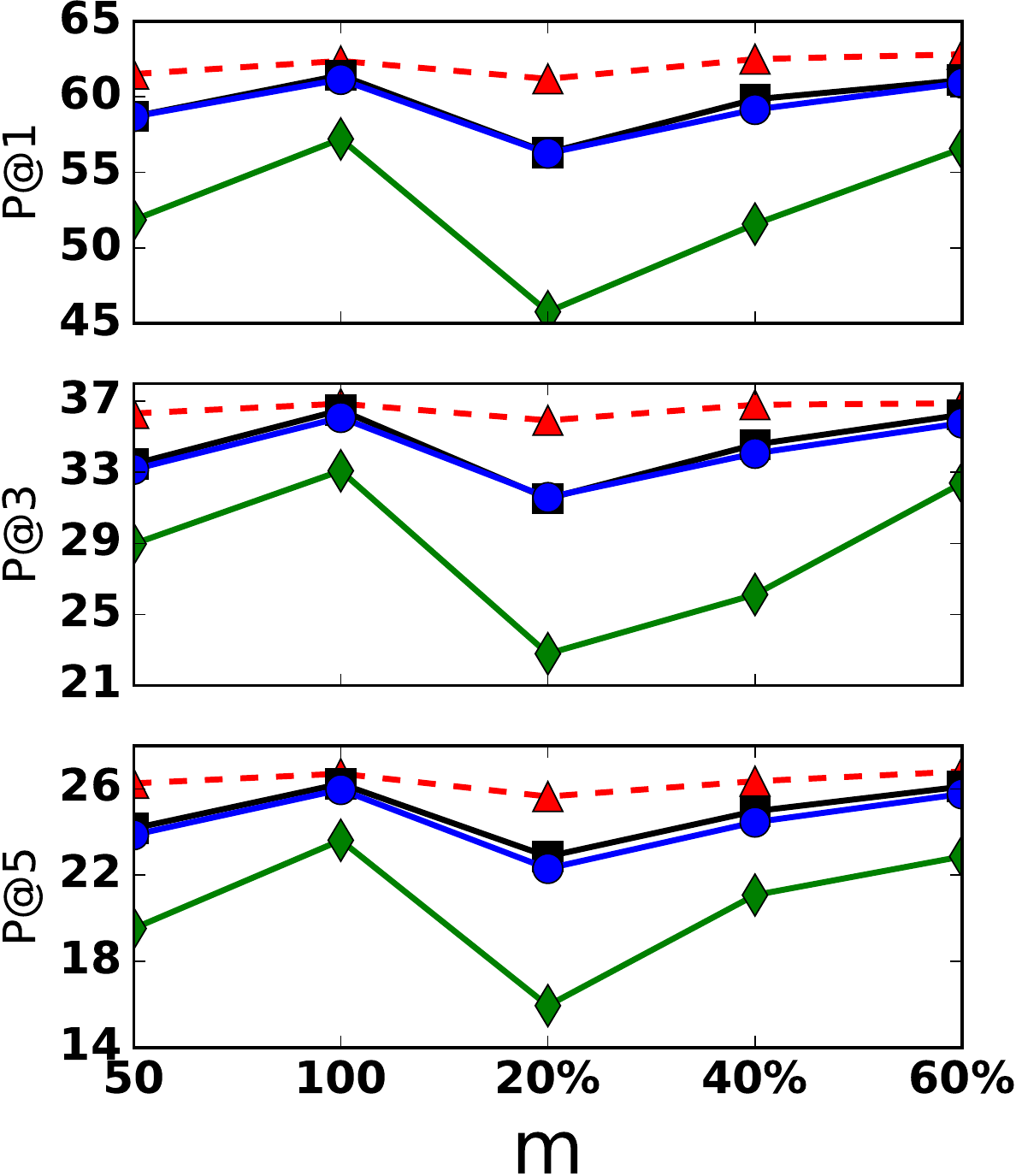}&
\includegraphics[scale=0.35]{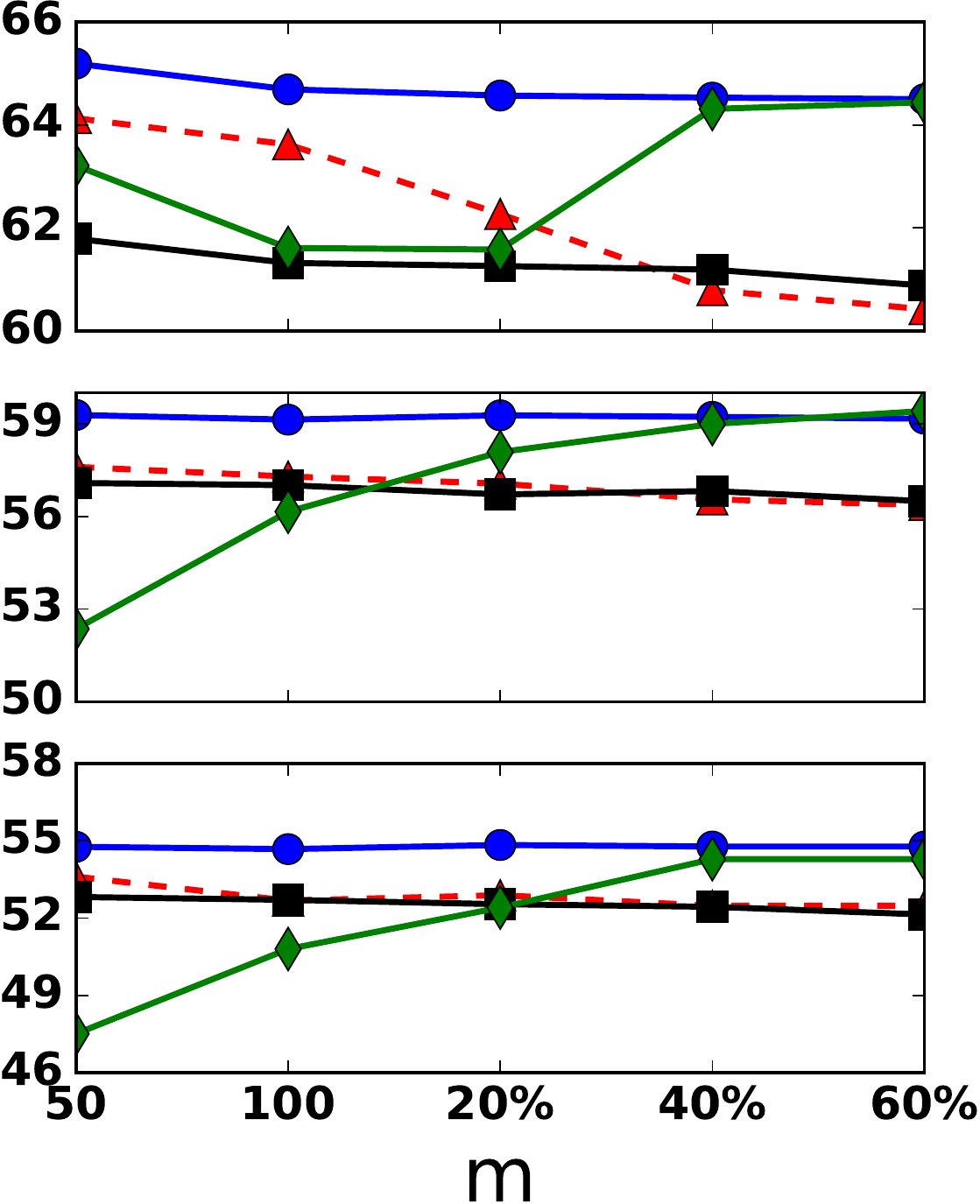}&
\includegraphics[scale=0.35]{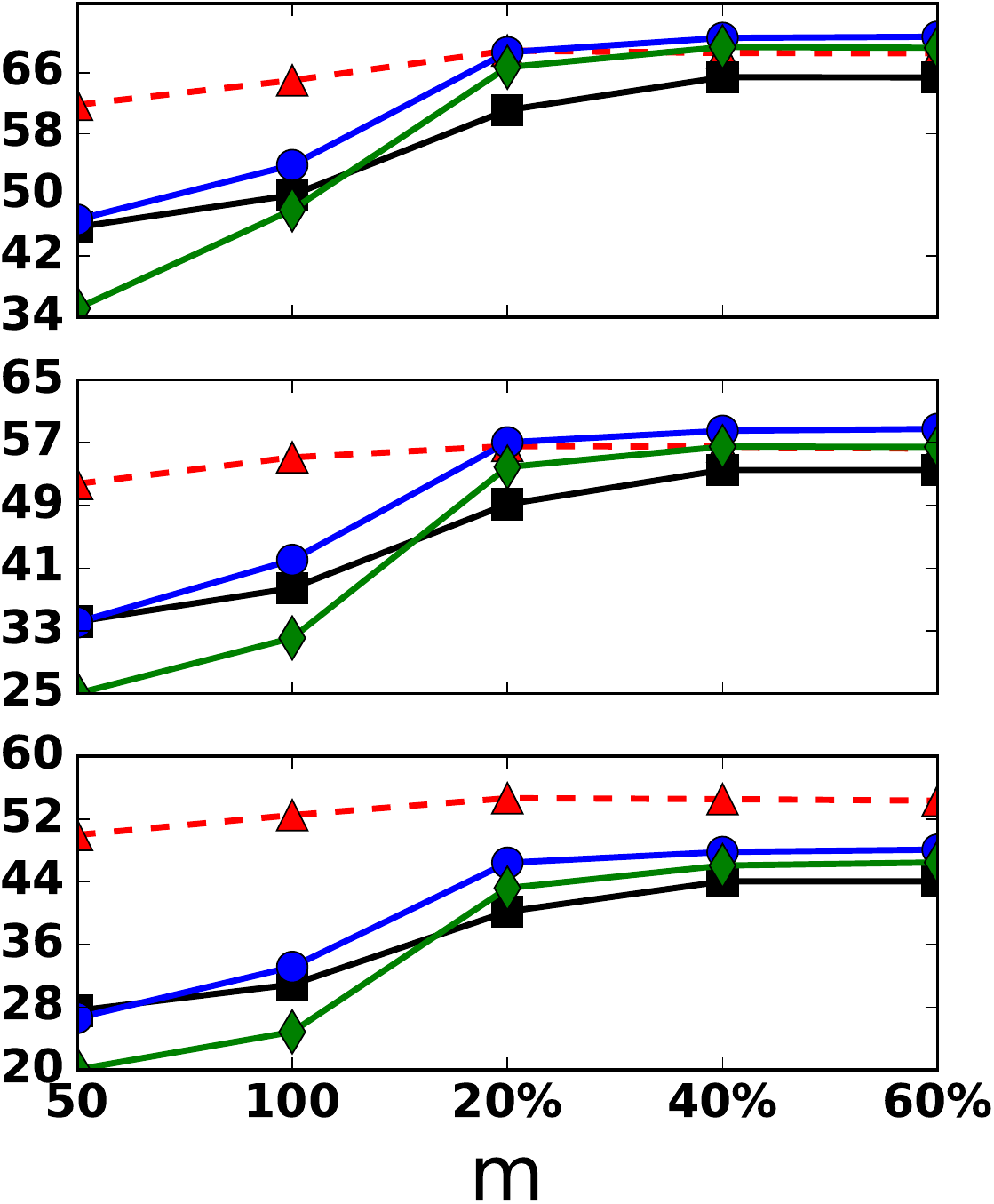}
\end{tabular}
\caption{Column 1, 2 and 3 represent Bibtex, Delicious, and EURLex datasets. Results for standard datasets: Bibtex, Delicious and EURLex. Legend: \textcolor{red}{RIPML (- -$\blacktriangle$- -)}, \textcolor{black}{LEML (--$\blacksquare$--)}, \textcolor{blue}{CPLST (--$\bullet$--)}, \textcolor{green}{CSSP (--$\blacklozenge$--)}. Y-axis: m = [50,~100,~ 20\% of $L$,~ 40\% of $L$,~ 80\% of $L$]. Here $k = 5$.}
\label{table:standard-datasets}
\end{figure*}

\begin{figure*}[h]
\centering
\begin{tabular}{ccc}
\includegraphics[scale=0.18]{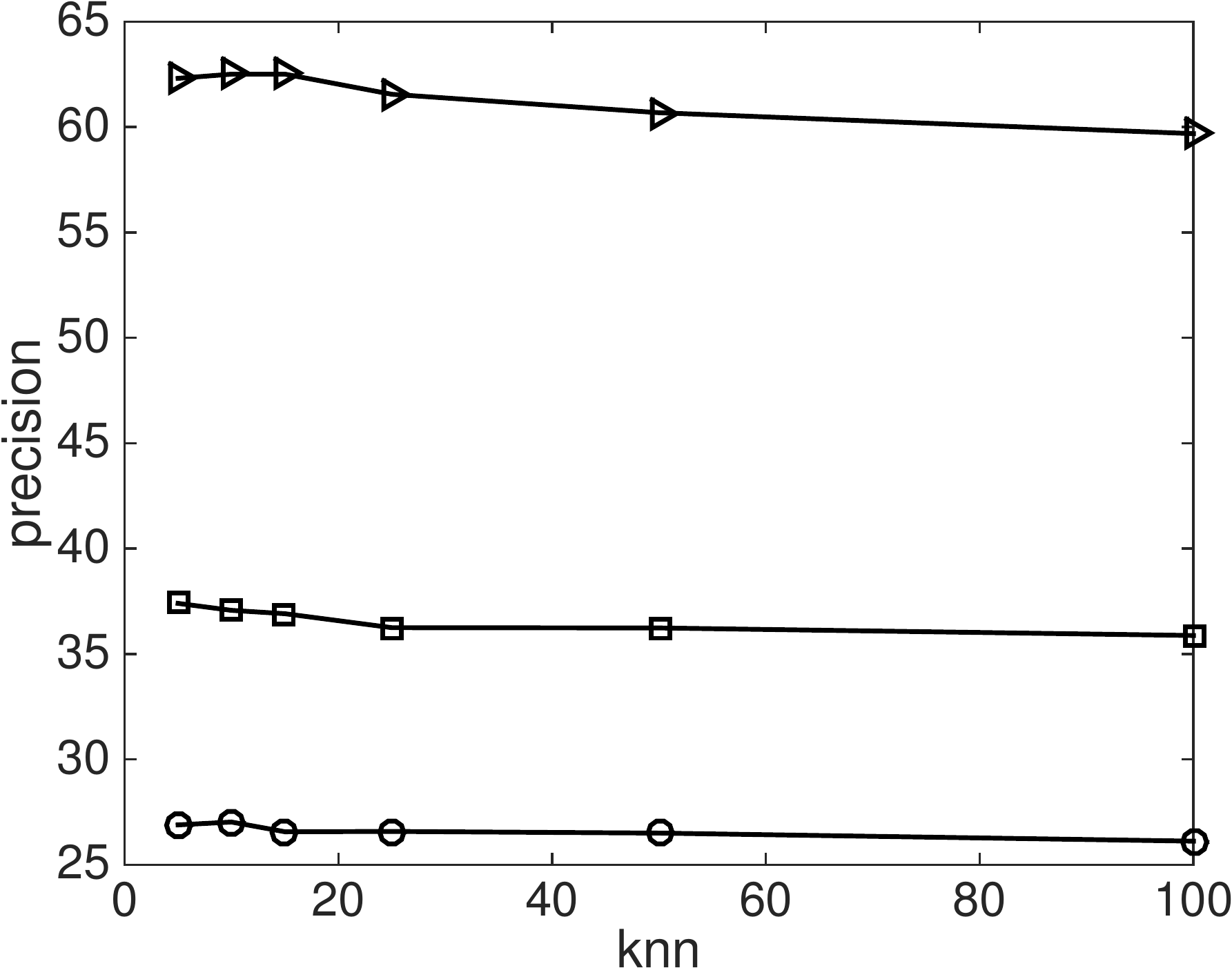}&
\includegraphics[scale=0.18]{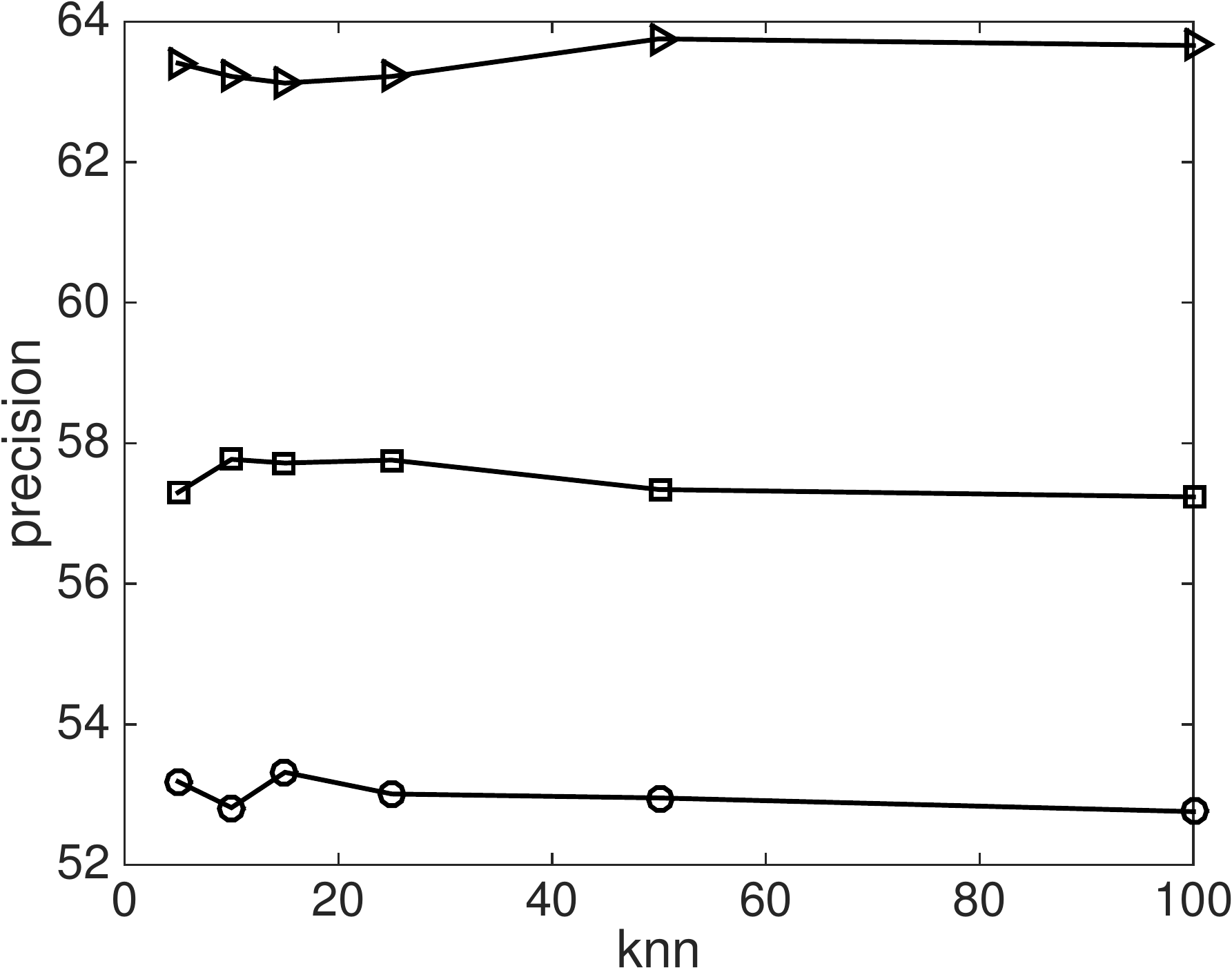}&
\includegraphics[scale=0.18]{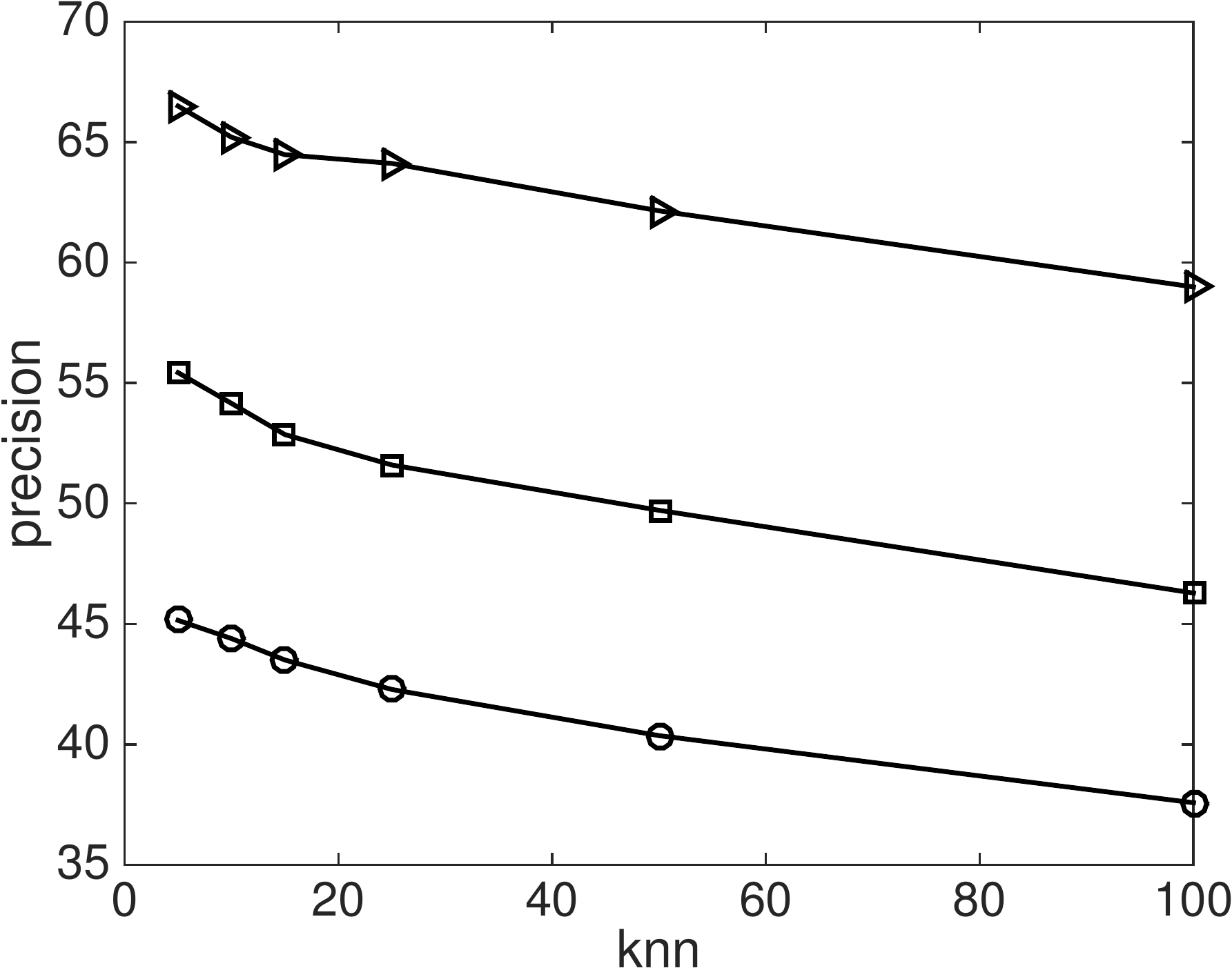}
\end{tabular}
\caption{Column 1, 2 and 3 represent Bibtex, Delicious, and EURLex datasets. Row-1: precision@\{1,3,5\} vs. k for kNN. 
Here --$\triangleright$--, --$\square$-- and --$\circ$-- corresponds to precision@1, 3 and 5 respectively. Here $m = 100$.}
\label{figure:standard-datasets}
\end{figure*}

\begin{figure*}[h]
\centering
\begin{tabular}{ccc}
\includegraphics[scale=0.35]{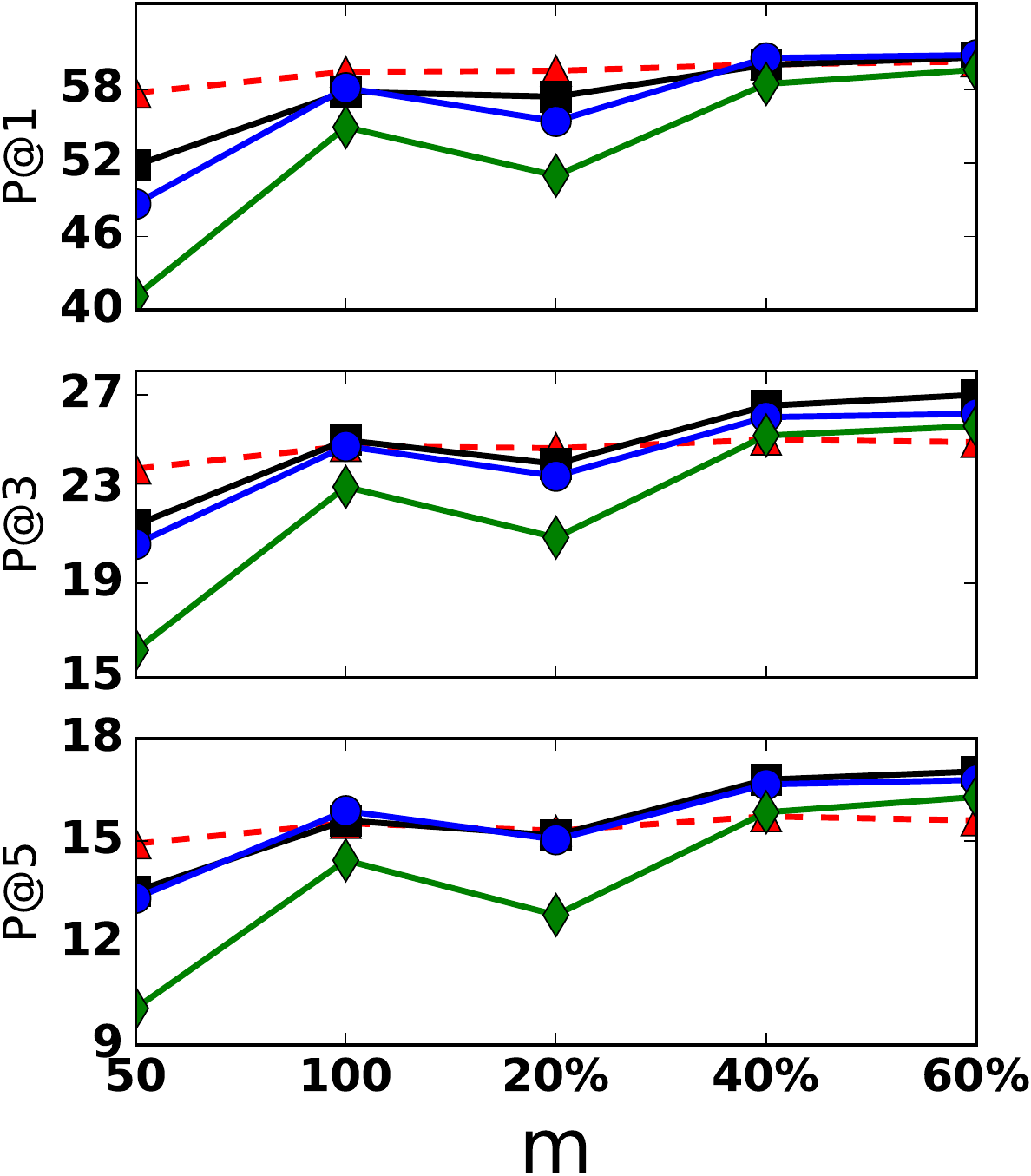}&
\includegraphics[scale=0.35]{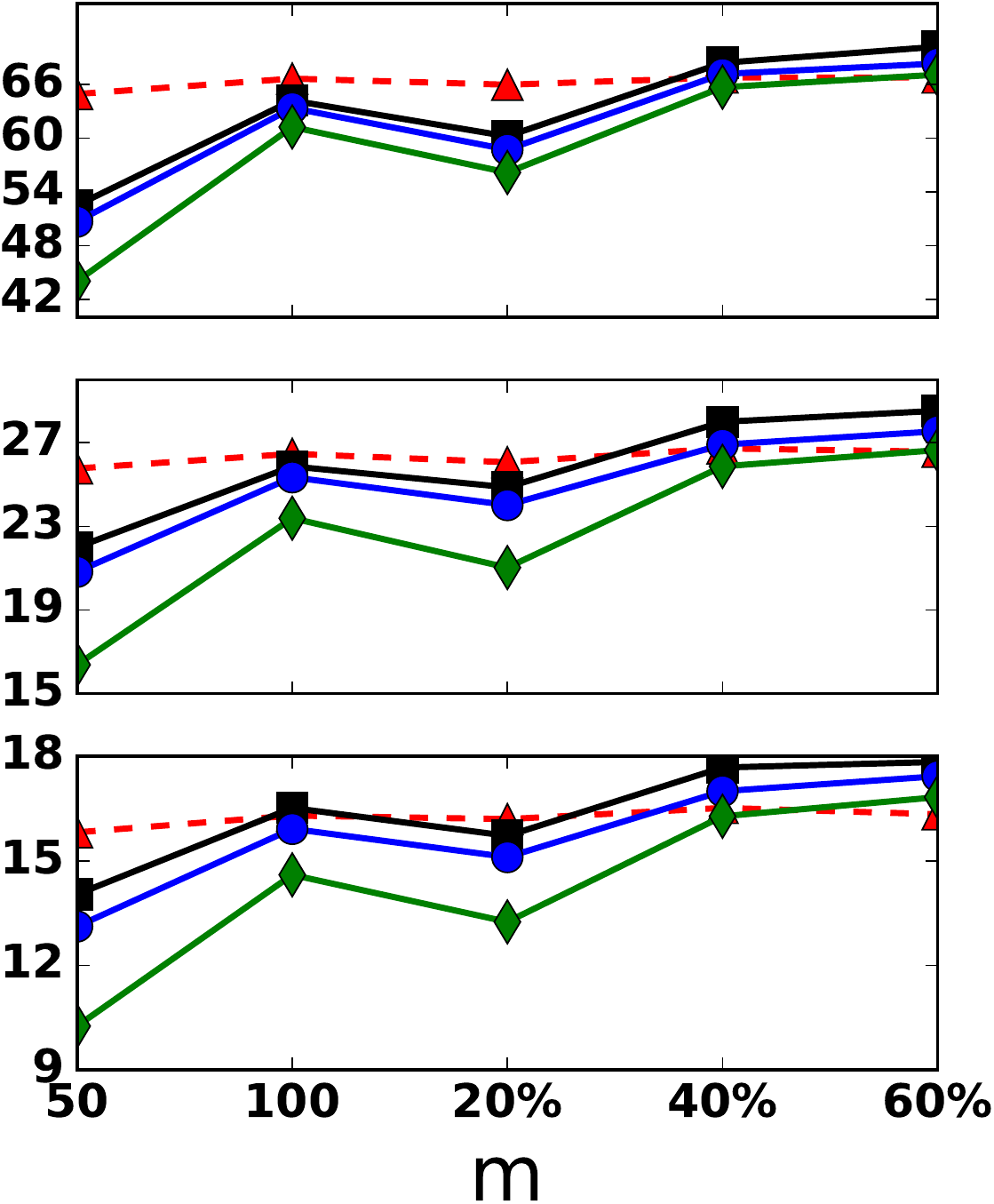}&
\includegraphics[scale=0.35]{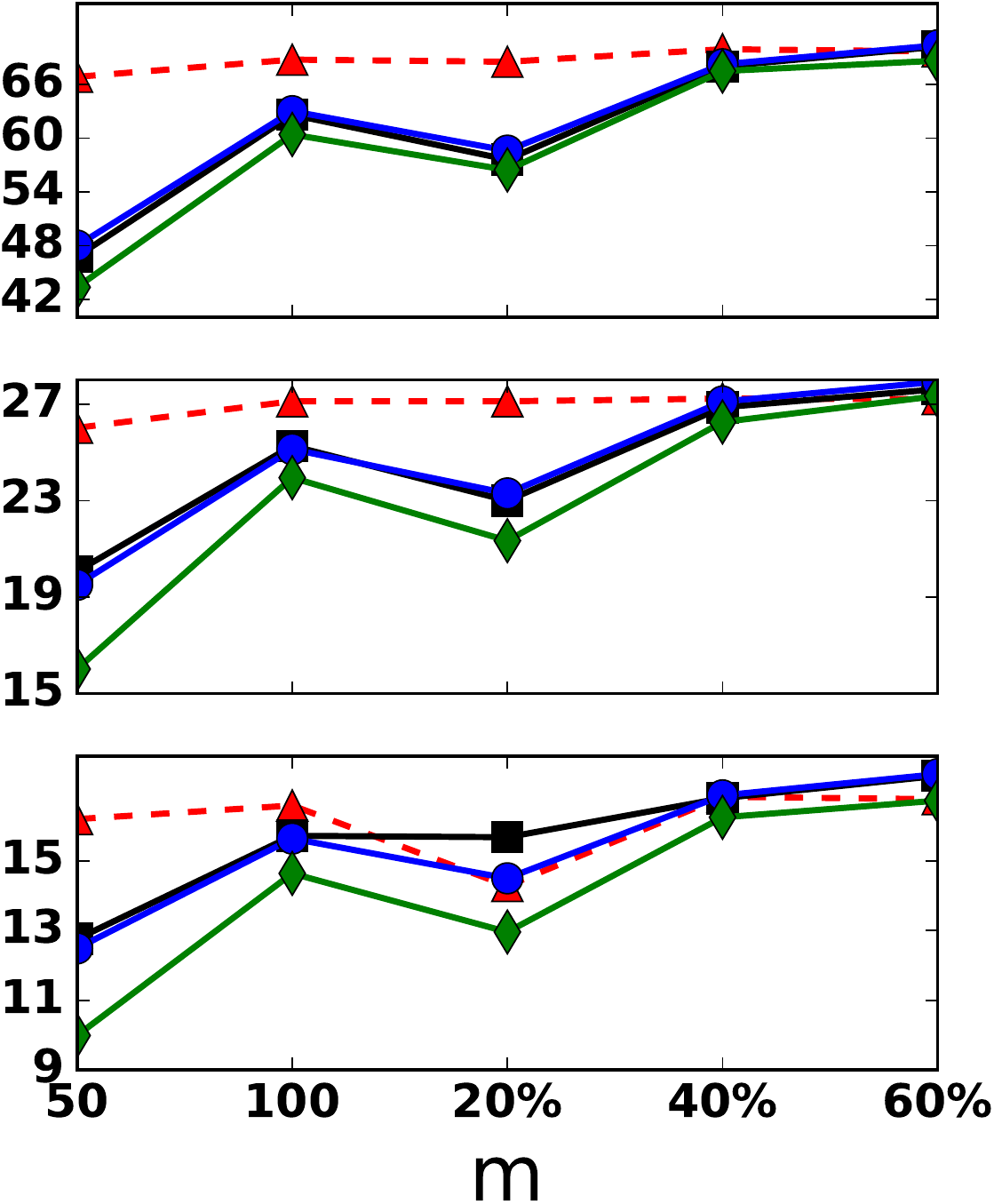}
\end{tabular}
\caption{Column-1, 2 and 3 correspond to $d = 200, 300,$ and $400$ respectively. Results on Chinese Relevance Modeling for different ambient dimensions. Legend: \textcolor{red}{RIPML (- -$\blacktriangle$- -)}, \textcolor{black}{LEML (--$\blacksquare$--)}, \textcolor{blue}{CPLST (--$\bullet$--)}, \textcolor{green}{CSSP (--$\blacklozenge$--)}. X-axis: m = [50,~100,~ 20\% of $L$,~ 40\% of $L$,~ 80\% of $L$]. Here $k = 5$.}
\label{table:ticker-cn}
\end{figure*}


\begin{table*}[ht]
\begin{center}
\small
\begin{tabular}{|c|p{0.8cm}|p{0.8cm}|p{0.8cm}|p{0.8cm}|p{0.8cm}|p{0.8cm}|p{0.8cm}|p{0.8cm}|p{0.8cm}|p{0.8cm}|p{0.8cm}|p{0.8cm}|}
 \hline
 \multicolumn{13}{|c|}{\bf Entity Recommendation (d = 400)}\\
 \hline
 & \multicolumn{3}{|c|}{\bf Clusters  = 1}  & \multicolumn{3}{|c|}{\bf Clusters  = 25}  &  \multicolumn{3}{|c|}{\bf Clusters = 43} &  \multicolumn{3}{|c|}{\bf Clusters = 56}\\
 \hline
  \small{m} & \small{P@1} & \small{P@3} & \small{P@5} & \small{P@1} & \small{P@3} & \small{P@5} &  \small{P@1} & \small{P@3} & \small{P@5} &  \small{P@1} & \small{P@3} & \small{P@5}\\
 \hline
 50   	& 	31.02 	&	23.60	&	20.20	&	 39.42  	&	31.72	&	27.61	& 	41.72	& 33.67	&	29.32	& 	41.23	& 	33.69	&	29.54 \\
 \hline 
  100   	& 	33.55 	&	25.58	&	21.70	&	 42.10  	&	34.24	&	29.89	& 	44.11	& 36.08	&	31.39	& 	43.91	& 	36.16	& 31.65\\
 \hline
   250   	& 	 35.76	&	27.20	&	23.46	&	 44.44  	&	35.81	&	31.38	& 	46.28	& 37.95	&	33.15	& 	46.30		& 	37.92	& 	33.45\\
 \hline
\end{tabular}
\caption{Results on Entity Recommendation dataset. Clusters = 1 means that clustering step was not done.}
\label{table:wiki}
\end{center}
\end{table*}

\begin{figure*}[h!]
\centering
\begin{tabular}{ccc}
\includegraphics[scale=0.18]{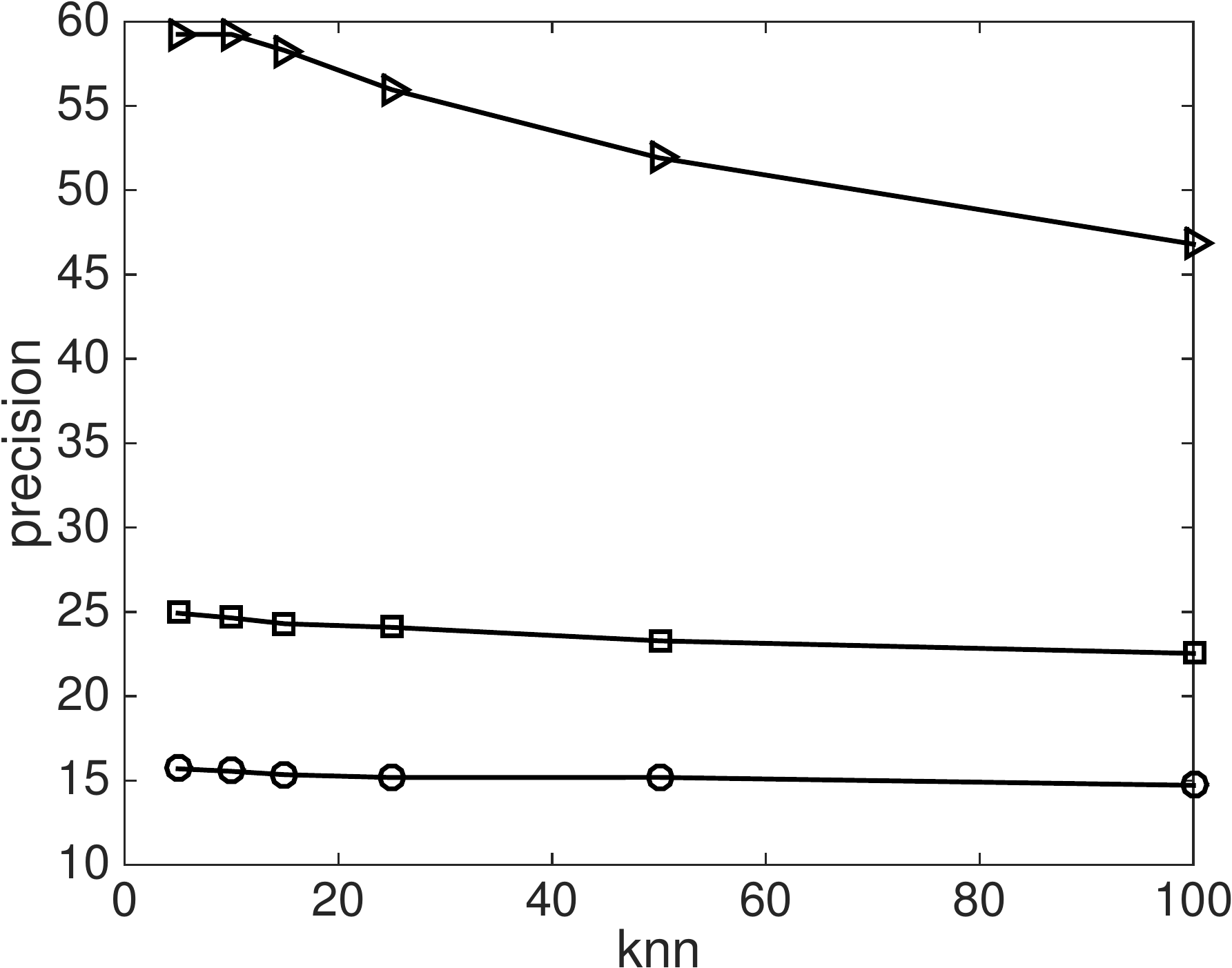}&
\includegraphics[scale=0.18]{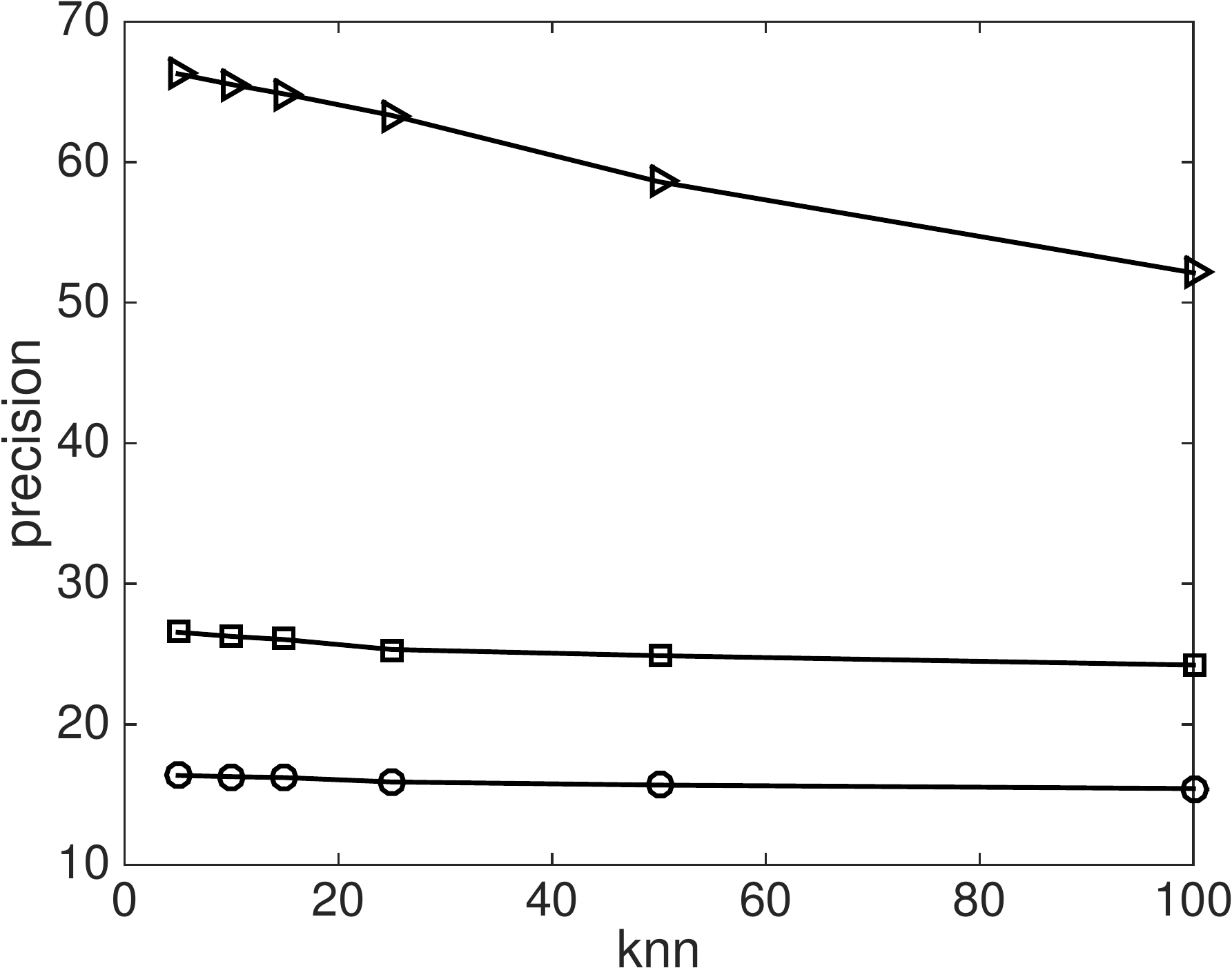}&
\includegraphics[scale=0.18]{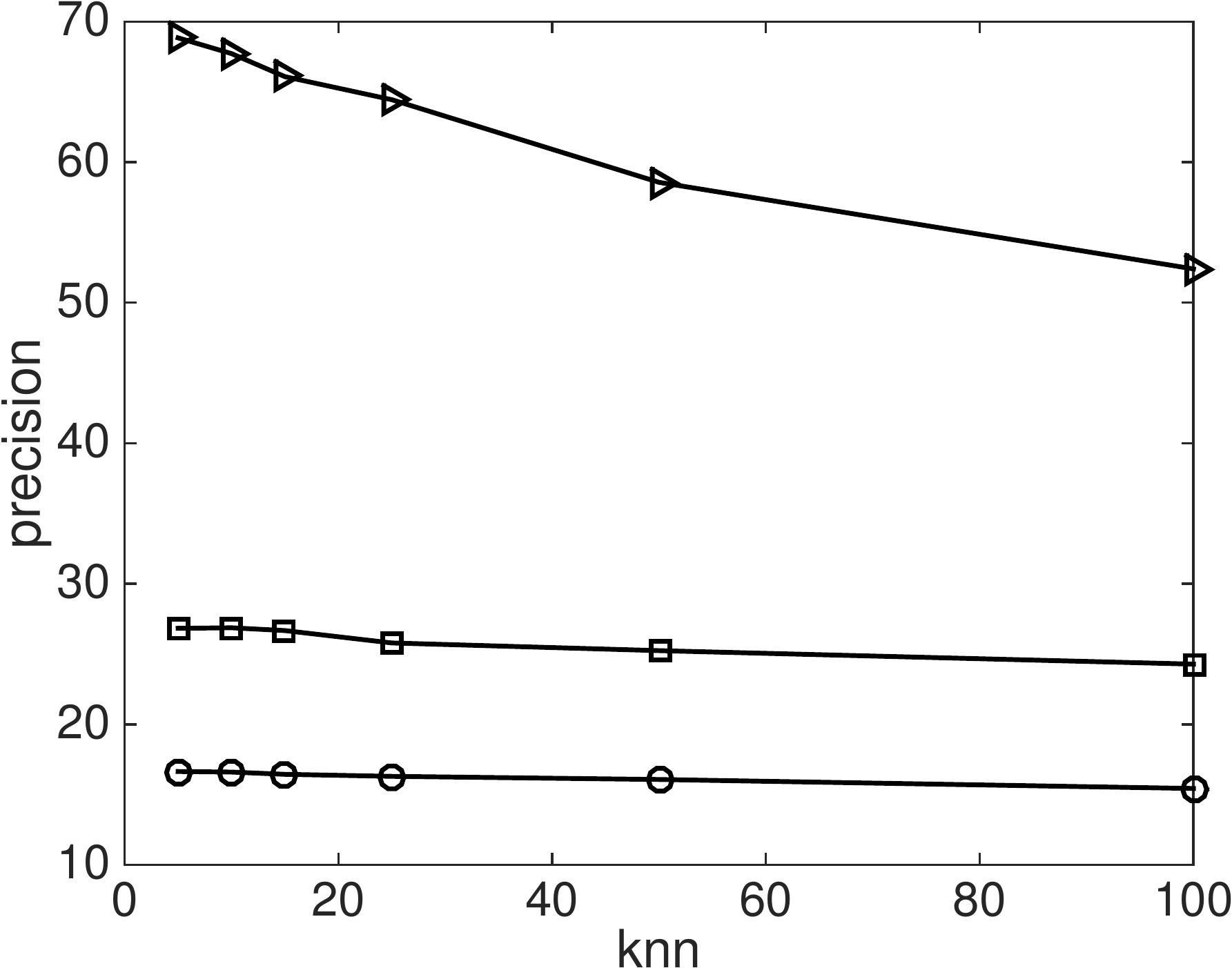}
\end{tabular}
\caption{Column-1, 2 and 3 correspond to $d = 200, 300,$ and $400$ respectively. Row-1: precision@\{1,3,5\} vs. k for kNN for Chinese Ticker dataset
Here --$\triangleright$--, --$\square$-- and --$\circ$-- corresponds to precision@1, 3 and 5 respectively. Here $m = 100$.}
\label{figure:ticker-kNN}
\end{figure*}

\section{Experiments and Results}
\subsection{Experimental Settings}

\noindent\textbf{Baselines:} Since our approach is based on linear dimensionality-reduction, we compare it with other state-of-the-art linear dimensionality reduction based approaches:
\begin{itemize}
\item LEML (Low rank Empirical risk minimization for Multi-Label Learning) with squared loss \cite{Yu:2014}. The implementation of this algorithm was provided by the authors. 
\item CPLST\footnote{The implementation for CPLST and CSSP was taken from: \url{https://github.com/hsuantien/mlc_lsdr}} (Conditional Principal Label Space Transformation) \cite{Chen:2012}. 
\item CSSP (Column Subset Selection Problem) \cite{Kwok:2013}
\end{itemize}


\noindent\textbf{Datasets:}

We  perform  experiments  on  five  textual real  world  datasets. The first three are popular datasets that have been used in the previous works: Bibtex \cite{Katakis:2008}, EURLex \cite{Loza:2010}, and Delicious \cite{Tsoumakas:2008}\footnote{All of these standard datasets are available online at: \url{http://mulan.sourceforge.net/datasets-mlc.html}}. These datasets are already partitioned into train and test which we use directly for our experiments.  We call these group of datasets as \textbf{MLL} datasets.

In order to further prove the generalizability and scalability of our approach, we conduct the same set of experiments on other datasets. These datasets were created and will be released as a by-product of our contribution to this work:   

\textbf{-Relevance Modeling (Chinese Finance News)}: a set of financial news documents in Chinese with their relevant ticker symbols. In this dataset, each document is labeled with ticker symbols of the companies to which the document is relevant. The annotation was performed by an expert native language editorial team. The evaluation results over these two datasets measure how well our approach deals with documents from other languages and emphasize the generalizability of our approach. 

\textbf{-Entity Recommendation (English Wikipedia)}: we randomly sampled one million documents from English Wikipedia and labeled the documents with entities. The concept of entity in this work is referred to any segment of text that is linked to another page in Wikipedia. We then filtered out the entities which occurred less than $10$ times in the entire dataset. Multilabel learning over such dataset is very challenging due to the number of labels (i.e., entities) in Wikipedia. This dataset measures how well our approach deals with very large number of labels and emphasize the scalability of our approach.

For the document embeddings of the above mentioned two datasets, we use doc2vec \cite{icml2014c2_le14} to learn  dense low-dimensional vectors. We train the embeddings of the words in documents using skip-bigram model \cite{Mikolov2013} using hierarchical softmax training. For the embedding of documents we exploit the distributed memory model since it usually performs well for most tasks \cite{icml2014c2_le14}. 

The dimension, number of labels and other statistics for the datasets are shown in Table \ref{table:datastats}. \\

\noindent\textbf{Evaluation Criteria:} Following the trail of the research in this field \cite{Yu:2014,Chen:2012,Bhatia:2015}, we take precision at $K$ (P@K) as our evaluation criteria. Precision at $K$ is the fraction of correct labels in the top-$K$ label predictions. 
For the ease of comparison with other research papers we use $K = 1,~3$ and $5$ in this paper.


\subsection{Results on MLL datasets}

The P@1, P@3 and P@5 results are tabulated in Figure \ref{table:standard-datasets}. We can observe that our results outperform the strong baselines in Bibtex and EURLex datasets. However, for Delicious dataset, our approach performs worse than CPLST. We suspect this might be due to the fact that the average number of nonzero labels per data-point is quite large for this dataset and also the features are very sparse, see Table \ref{table:datastats}.

An interesting point to note from these results is that RIPML remains stable with varying $m$ while other approaches start with a lower precision for small $m$ and then gradually improve with increasing $m$. This stability of RIPML can be attributed to RIP which allows to obtain a stable distance-preserving embedding with $m = \cO(s\log(s))$ where $s$ is the maximum number of nonzeros in the label vectors. Thus, increasing $m$ above a certain threshold only results in a marginal improvement. This property also allows RIPML to perform better than other algorithms for small $m$ -- see results for Bibtex, and EURLes for $m = 50$ or $100$. 

Figure \ref{figure:standard-datasets} shows the variation of precision with number of nearest neighbors used for kNN. Bibtex and Delicious are quite stable with the choice of number of nearest neighbors but EURLex performs better with smaller number of nearest neighbors.  

\subsection{Results on Relevance Modelling Datasets}
In  this  work,  we  obtain the low-dimensional embedding of documents using the method described in Section 3 with window sizes of 10.   In order to see the effect of dimensions, we  also  experiment  with  four low-dimensional models (200, 300 and 400 dims).  During training the document embeddings we limit the number of iterations to 10 to increase the efficiency. 

Figure \ref{table:ticker-cn} shows the results for Chinese relevance modeling datasets with different ambient dimensions. All these results are averaged over $5$ random train-test splits. It is interesting to note that RIPML performs better than CPLST and CSSP always, but performs worse than LEML in certain cases. Another interesting observation is that even though the P@1 is good for these datasets, P@5 drops significantly for all the approaches. This is due to the very low number of relevant tickers for each document (average number of relevant tickers per document is about $1$). In conclusion, RIPML performs well for detecting the relevant tickers, considering that our approach doesn't require any linguistic preprocessing which is one of the main challenges in multilingual NLP community. 

Figure \ref{figure:ticker-kNN} shows the variation of precision with respect to number of nearest neighbors used during prediction for these two datasets. For these datasets, small number of nearest neighbors results in a better precision, due to very less average number of relevant ticker symbols per document.

\subsection{Results on Entity Recommendation Dataset}

We created this dataset from Wikipedia to show how clustering can be used to scale to big datasets with very large number of labels, and the effect of clustering on performance. In order to get a baseline to compare the effect of clustering, we first conduct experiments without using the clustering step. Given that our training and testing procedure is very efficient, we were able to train our model on this data in less than $3$ minutes on a laptop\footnote{Apple MacBook Pro with 2.5GHz Intel Core i7 and 16 GB RAM.}. We used $10000$ data points for testing and rest for training. Our results are averaged over $5$ random train-test split. It took approximately $72$ milliseconds to predict labels for each test data point. We are not providing any comparison for this dataset with other approaches because we were not able to train a model within reasonable amount of time. 

In order to improve the performance for this challenging dataset, we then applied kMeans clustering algorithm to cluster the training data (features) into different number of clusters and trained RIPML on each of the clusters separately by borrowing ideas from \cite{Bhatia:2015}. For every test data-point, we first figure out which cluster it belongs to by finding the nearest cluster center and then apply the kNN based prediction procedure on that cluster. Clustering helps us in two ways -- it makes prediction faster by allowing us to do kNN on a small amount of data and it also increases the prediction accuracy. This is evident from the results shown in Table \ref{table:wiki} where we present precision for varying number of clusters. Going from no clustering to around $43$ clusters increases the precision by about $10\%$. With $43$ clusters it took $\sim 7$ minutes to train and approximately $6.5$ milliseconds to predict labels for each test data point.

\section{Conclusions}
In this paper, we presented a novel, scalable, and general multilabel learning algorithm based on random-projections and kNN called RIPML. We demonstrated its performance on six different real world datasets which includes three popular and three new datasets. The new datasets would also be released for the use of research community as a part of this work.

We would like to extend our algorithm to explore the problem of missing label cases. We also plan to study the impact of using different RIP, satisfying random/deterministic ensembles. Moreover we will investigate the performance of other loss functions for the regression phase.

\bibliographystyle{ieeetr}
\bibliography{acl2016}

\begin{thebibliography}{10}

\bibitem{Qi:2007}
G.-J. Qi, X.-S. Hua, Y.~Rui, J.~Tang, T.~Mei, and H.-J. Zhang, ``Correlative
  multi-label video annotation,'' in {\em Proceedings of the 15th ACM
  International Conference on Multimedia}, MM '07, (New York, NY, USA),
  pp.~17--26, ACM, 2007.

\bibitem{Barutcuoglu:2006}
Z.~Barutcuoglu, R.~E. Schapire, and O.~G. Troyanskaya, ``Hierarchical
  multi-label prediction of gene function,'' {\em Bioinformatics}, vol.~22,
  pp.~830--836, Apr. 2006.

\bibitem{Katakis:2008}
I.~Katakis, G.~Tsoumakas, and I.~Vlahavas, ``Multilabel text classification for
  automated tag suggestion,'' in {\em In: Proceedings of the ECML/PKDD-08
  Workshop on Discovery Challenge}, 2008.

\bibitem{Boutell:2004}
M.~R. Boutell, J.~Luo, X.~Shen, and C.~M. Brown, ``Learning multi-label scene
  classification,'' {\em Pattern Recognition}, vol.~37, no.~9, pp.~1757 --
  1771, 2004.

\bibitem{Agrawal:2013}
R.~Agrawal, A.~Gupta, Y.~Prabhu, and M.~Varma, ``Multi-label learning with
  millions of labels: Recommending advertiser bid phrases for web pages,'' in
  {\em Proceedings of the 22Nd International Conference on World Wide Web}, WWW
  '13, (New York, NY, USA), pp.~13--24, ACM, 2013.

\bibitem{Hsu:2009}
D.~Hsu, S.~M. Kakade, J.~Langford, and T.~Zhang, ``Multi-label prediction via
  compressed sensing,'' {\em CoRR}, vol.~abs/0902.1284, 2009.

\bibitem{Kapoor:2012}
A.~Kapoor, R.~Viswanathan, and P.~Jain, ``Multilabel classification using
  bayesian compressed sensing,'' pp.~2645--2653, 2012.

\bibitem{Yu:2014}
H.-F. Yu, P.~Jain, P.~Kar, and I.~S. Dhillon, ``{Large-scale Multi-label
  Learning with Missing Labels},'' in {\em ICML}, 2014.

\bibitem{Bhatia:2015}
K.~Bhatia, H.~Jain, P.~Kar, P.~Jain, and M.~Varma, ``Locally non-linear
  embeddings for extreme multi-label learning,'' {\em CoRR},
  vol.~abs/1507.02743, 2015.

\bibitem{Rudelson:2006}
M.~Rudelson and R.~Vershynin, ``Sparse reconstruction by convex relaxation:
  Fourier and gaussian measurements,'' in {\em Information Sciences and
  Systems, 2006 40th Annual Conference on}, pp.~207--212, IEEE, 2006.

\bibitem{Dasgupta:2003:JL}
S.~Dasgupta and A.~Gupta, ``An elementary proof of a theorem of johnson and
  lindenstrauss,'' {\em Random Structures \& Algorithms}, vol.~22, no.~1,
  pp.~60--65, 2003.

\bibitem{Indyk:1998:knn}
P.~Indyk and R.~Motwani, ``Approximate nearest neighbors: towards removing the
  curse of dimensionality,'' in {\em Proceedings of the thirtieth annual ACM
  symposium on Theory of computing}, pp.~604--613, ACM, 1998.

\bibitem{Datar:2004:locality}
M.~Datar, N.~Immorlica, P.~Indyk, and V.~Mirrokni, ``Locality-sensitive hashing
  scheme based on p-stable distributions,'' in {\em Proceedings of the
  twentieth annual symposium on Computational geometry}, pp.~253--262, ACM,
  2004.

\bibitem{Chen:2012}
Y.-n. Chen and H.-t. Lin, ``Feature-aware label space dimension reduction for
  multi-label classification,'' in {\em Advances in Neural Information
  Processing Systems 25} (F.~Pereira, C.~J.~C. Burges, L.~Bottou, and K.~Q.
  Weinberger, eds.), pp.~1529--1537, Curran Associates, Inc., 2012.

\bibitem{Kwok:2013}
W.~Bi and J.~Kwok, ``Efficient multi-label classification with many labels,''
  in {\em Proceedings of the 30th International Conference on Machine Learning
  (ICML-13)} (S.~Dasgupta and D.~Mcallester, eds.), vol.~28, pp.~405--413, JMLR
  Workshop and Conference Proceedings, May 2013.

\bibitem{Loza:2010}
E.~Loza~Menc{\'{\i}}a and J.~F{\"{u}}rnkranz, ``Efficient multilabel
  classification algorithms for large-scale problems in the legal domain,'' in
  {\em Semantic Processing of Legal Texts -- Where the Language of Law Meets
  the Law of Language} (E.~Francesconi, S.~Montemagni, W.~Peters, and
  D.~Tiscornia, eds.), vol.~6036 of {\em Lecture Notes in Artificial
  Intelligence}, pp.~192--215, Springer-Verlag, 1~ed., May 2010.
\newblock accompanying EUR-Lex dataset available at
  \url{http://www.ke.tu-darmstadt.de/resources/eurlex}.

\bibitem{Tsoumakas:2008}
G.~Tsoumakas, I.~Katakis, and I.~Vlahavas, ``{Effective and Efficient
  Multilabel Classification in Domains with Large Number of Labels},'' in {\em
  Proc. ECML/PKDD 2008 Workshop on Mining Multidimensional Data (MMD'08)},
  p.~XX, 2008.

\bibitem{icml2014c2_le14}
Q.~Le and T.~Mikolov, ``Distributed representations of sentences and
  documents,'' in {\em Proceedings of the 31st International Conference on
  Machine Learning (ICML-14)} (T.~Jebara and E.~P. Xing, eds.), pp.~1188--1196,
  JMLR Workshop and Conference Proceedings, 2014.

\bibitem{Mikolov2013}
T.~Mikolov, K.~Chen, G.~Corrado, and J.~Dean, ``Efficient estimation of word
  representations in vector space,'' {\em CoRR}, vol.~abs/1301.3781, 2013.

\end{thebibliography}
\end{document}